\documentclass[twocolumn,prX]{revtex4}
\usepackage{graphicx}


\begin{document}

\title{A fractal approach to the rheology of concentrated cell suspensions}

\author{A. Iordan$^{a}$, A. Duperray$^{b,c}$, C. Verdier$^{a}$%
%\footnote{Author to whom correspondence should be addressed verdier@ujf-grenoble.fr
%Tel. +33 4 76 63 59 80 
%Fax +33 4 76 63 54 95} 
}

\affiliation{(a) Laboratoire de Spectrom\'etrie Physique, CNRS and Universit\'e Joseph-Fourier (UMR5588), 
140 avenue de la physique, BP87 -- 38402 Saint Martin d'H\`eres cedex, France.
(b) INSERM, U823, Grenoble, France.
(c) Universit\'e Grenoble I, Institut Albert Bonniot, Centre de Recherche Ontog\'en\`ese -- 
Oncog\'en\`ese mol\'eculaires, Grenoble, France.
}

\begin{abstract}
Results on the rheological behavior of novel CHO cell suspensions in a large range of concentrations are reported. 
The concentration--dependent yield stress and elastic plateau modulus are formalized in the context of fractal aggregates under shear, and 
quite different exponents are found as compared to the case of red blood cell suspensions. This is explained in terms of intrinsic 
microscopic parameters such as the cell-cell adhesion energy and cell elasticity but also the cell's individual dynamic properties, 
found to correlate well with viscoelastic data at large concentrations ($\phi \geq 0.5$).
\end{abstract}


\maketitle

The rheology of complex fluids has been studied extensively over the past decades \cite{Larson1999} and has revealed very intriguing
behaviors, in particular properties of suspensions, either micronic or colloidal, are still a subject of interest 
\cite{coussot2005,flatt2006,Alexandrou2007,Brader2007}. Classical behaviors of suspensions reveal shear--thinning effects usually, but other unusual
ones like shear--thickening \cite{Laun1984} (i.e. viscosity increase with shear rate) or yield stress have been observed 
\cite{coussot2005,flatt2006}. The yield stress is the critical value of the shear stress needed to induce flow for a given fluid. It is closely related to 
the internal structure of the fluid therefore its ability to form (or break) particle clusters under flow. In this respect many studies have focused on
solid sphere suspensions.

On the other hand, there are much less works dedicated to suspensions of deformable particles, such as biological cell 
suspensions. The main works can be found in the field of blood rheology. Suspensions of Red Blood Cells (RBC) within plasma were first investigated 
by Chien \cite{chien1967,chien1967b} and revealed a shear--thinning behavior, but a more detailed inspection of the viscosity--shear rate 
diagrams showed that at low shear rates, the stress level is close to a constant $\sigma_{s}$ (Pa), called 
the yield stress. The well--known Casson's model \cite{Casson1959}, $\sqrt \sigma=\sqrt\sigma_{s}+\sqrt{\mu \dot\gamma}$, relating the shear 
stress $\sigma$ to the shear rate $\dot\gamma$ ($\mu$ being a constant viscosity) can be used to determine the yield stress. Chien and co--authors
obtained $\sigma_{s}$ for a large range of hematocrit (H), i.e. the RBC volume concentration \cite{Chien1966}. They showed a relationship of the 
type $\sigma_{s}\sim (H-b)^{3}$ ($b$ being a constant hematocrit) as also observed in a recent work \cite{Picart1998a}. 

It is still not known yet whether this type of behavior is universal, or rather it could depend on 
cell type, cell shape or other biological effects such as cell adhesion or cell elasticity. In particular, one proposed explanation of 
the yield stress in RBCs suspensions is based on the existence of "rouleaux" which build due to cell interactions and exhibit large shape 
aspect ratios \cite{chien1967b} and a fractal dimension $D$. Therefore it is necessary to 
apply strong enough stresses in order to break such aggregates, in close relation with the yield stress.

In this letter we propose to investigate the rheology of a new cell suspension, consisting of CHO cells (Chinese Hamster Ovary cells) in a large 
range of concentrations. Such cells are commonly used in biology, easy to culture, and can be genetically modified to induce different adhesive 
properties. These cells are spherical when suspended in a culture medium, and organized in a specific manner leading to 
particular aggregation patterns of fractal type. This leads to the determination of scaling laws based on fractal exponents 
(for the yield stress $\sigma_{s}$ and elastic modulus $G_{0}$) which are seen to be non universal but dependent on cell type. 
The flow curves constitute a basis to test classical empirical models (Bingham, Casson, Herschel--Bulkley models) and 
other ones \cite{snabre1996,snabre1999} based on kinetic theories describing the rupture and formation of particle clusters. The latter 
ones successfully relate macroscopic effects to microscopic parameters, such as the cell--cell adhesion energy and the cell elasticity. These microscopic 
parameters remarkedly match those found in the litterature using other techniques. This is important in the context of recent studies related 
to tumour growth \cite{Drasdo2005,galle2005,Ambrosi2007} which consider cell assemblies with interactions as well as cell elastic deformations. 
Furthermore, this study emphasizes the relationship between the dynamic rheological properties of suspensions \cite{Verdier2003} and the single cell 
properties.

{\it Experiments} -- In our model system, CHO cells are grown in culture medium using standard $T75$ boxes under proper conditions ($37^{\circ} C$, $5 \% \,CO_{2}$), 
until they are at $70\%$ confluence, then are detached using trypsin, are mixed together and centrifugated at $1200 \,rpm$, a high enough 
velocity to get a concentrated suspension, but slow enough in order to maintain the cells alive. Cell volume concentration $\phi$ 
(i.e. similar to the hematocrit $H$) is determined
accurately after centrifugation in hematocrit tubes containing the CHO cells. Then the right amount of remaining supernatant is removed until 
the desired concentration is obtained (between $0$ and $60\,\%$). Different experiments were carried out on a conventional Rheometer 
(Bohlin Gemini 150). Both steady shear and oscillatory measurements were made at $T=20^{o}C$. Due to the large amount of cells needed (we usually 
require twelve $T75$ flasks in order to obtain a volume of roughly $0.3\, mL$ of cells), we chose to use a plate--plate geometry 
($20\,mm$ diameter) with a small gap 
(between $400 \,\mu m$ and $1 \,mm$) for the concentrated suspensions whereas the smaller concentrations (below $10\,\%$) were tested using the 
$60 \,mm$ cone--plane geometry ($2^{o}$ angle). Typically in our fluid, the suspended cells are spherical and monodisperse with a radius 
$a \sim  10 \mu m$.

\begin{figure}
\includegraphics[width=0.9\columnwidth]{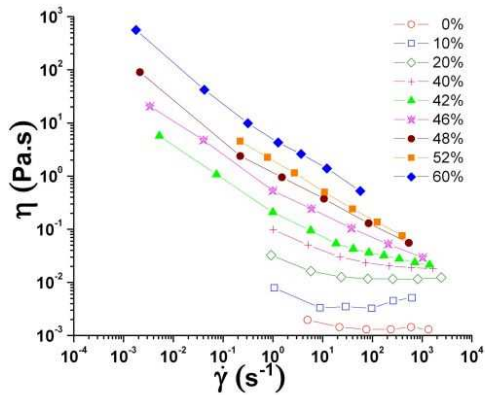}
\caption{\label{Fig:1} Viscosity $\eta$ (Pa.s) vs. shear rate $\dot\gamma$ ($s^{-1}$) at different volume concentrations $\phi$ from $0$ to $60\%$}
\end{figure}

{\it Results} -- Experimental results for constant steady state shear rate $\dot\gamma$ are presented
in Fig.\ref{Fig:1}. The viscosity $\eta$ is shown to vary over several decades, within shear rates typically between $10^{-3}s^{-1}$ and 
$10^{3}s^{-1}$. In some cases, we limited ourselves to the higher shear rates because of experimental reasons (i.e. steady state 
not reached). By a first inspection of the curves, we recognize the signature of a yield stress fluid as depicted by the slope close to $-1$ in the viscosity--shear rate 
diagram (or equivalently a constant shear stress at low shear rates), especially at the largest concentrations $\phi$, which will be particularly of interest here. 
The existence of this yield stress is attributed to weak interactions which can exist after preparation of the system. Already existing proteins 
are available on cell membranes and can be recruited to form bonds, leading to particular structure arrangments. This explains the presence 
of a yield stress related to the formation of such structures. The yield stress is found to depend on volume concentration $\phi$ in a manner to be discussed 
later.
 
\begin{figure}
\includegraphics[width=0.8\columnwidth]{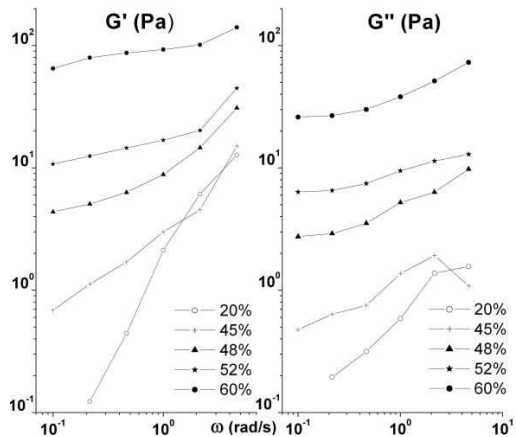}
\caption{\label{Fig:2} Shear moduli $G'$ and $G"$ (Pa) vs. angular frequency $\omega$ at different volume concentrations 
$\phi$ ranging from $20\%$ to $60\%$}
\end{figure}

A second series of experiments was carried out in order to study the systems under oscillatory strains at angular frequency $\omega$. Small 
deformations ($1\%$ or less) within the 
linear regime were performed in order to characterize the elastic modulus $G'(\omega)$ and the loss modulus $G"(\omega)$. We find an interesting 
behavior as shown in Fig.\ref{Fig:2}. Moduli $G"(\omega)$ usually prevails over $G'(\omega)$
at small concentrations (e.g. $\phi=0.2$), but as $\phi$ increases, the system becomes elastic with a much larger $G'(\omega)$. This behavior
is the signature of a viscoelastic medium, due to the fact that interactions between elastic cells become effective at large concentrations 
($\phi\geq 0.4$). The slow increase of the elastic modulus $G'$ against frequency reveals the presence of a so--called 'elastic plateau' 
modulus ($G_{0}$) determined by the value of $G'(\omega)$ at intermediate rates ($1\, rad/s$ typically). The presence of elasticity has been 
observed previously for RBC suspensions \cite{Thurston1972}, above a critical volume fraction around $\phi=0.2$, and is believed to come from 
the elasticity of the cells as they are packed more closely at large concentrations such as the ones also encountered 
in tumour spheroids \cite{galle2005}. Finally, we observe that the trends in the $G'-G"$ plots for large concentrations ($\phi \geq 0.5$) 
are remarkedly similar to previous microrheological results obtained on single cells \cite{Fabry2001,Alcaraz2003,Hoffman2006}.

As in the case of suspensions, we define a maximum packing fraction $\phi_{0}$ (which is usually $0.64$ or even $0.74$ for solid spheres in a 
face--centered--cubic crystal), depending on cell elasticity, i.e. their compactness \cite{quemada1998a}. Due to the presence of soft spherical cells, 
it is expected that the value of $\phi_{0}$ will be in this range. $\phi_{0}$ is determined using the reduced viscosity plot 
$\frac{\eta}{\eta_{0}}$ as a function of $\phi$ ($\eta$ at a shear rate of $10^{3} s^{-1}$, $\eta_{0}=0.0014 Pa.s$ the solvent viscosity). In our case, 
this data (not shown) is found to match the well--known 
equation $\frac{\eta}{\eta_{0}}=(1-\frac{\phi}{\phi_{0}})^{-2.5\phi_{0}}$ proposed earlier \cite{Krieger1959}, this providing the 
value $\phi_{0}\sim 0.65$.

\begin{figure}
\includegraphics[width=0.9\columnwidth]{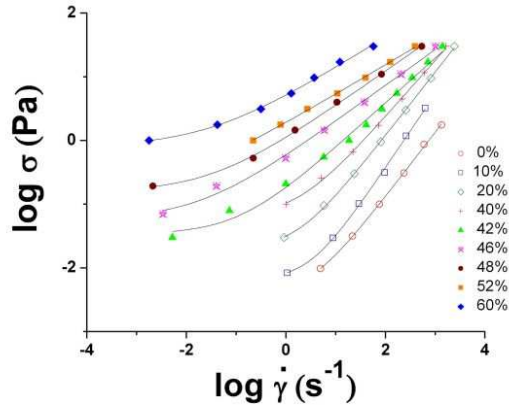}
\caption{\label{Fig:3} Determination of the yield stress $\sigma_{s}$ using Herschel--Bulkley's model.}
\end{figure}

In order to investigate the effect of the volume concentration $\phi$, we first need to obtain the flow curve $\sigma(\dot \gamma)$ of the suspensions, 
as well as the relevant parameters, such as the yield stress $\sigma_{s}$. From the viscosity curve in Fig.\ref{Fig:1}, we plot the stress 
$\sigma = \eta(\dot \gamma) \dot \gamma$ 
vs. shear rate $\dot \gamma$ in Fig.\ref{Fig:3}, and fit the data with the Herschel--Bulkley law \cite{coussot2005} $\sigma = \sigma_{s} + K \dot \gamma ^{n}$, 
where $K$ is a constant, and $n$ is a shear--thinning exponent ranging between $0$ and $1$ ($1$ is for a 
Bingham fluid, and the case of the Newtonian fluid is recovered for $n=1$, $\sigma_{s}=0$). Parameters have been optimized using a standard 
Newton--Raphson method.

This leads to the determination of the yield stress $\sigma_{s}$ as a function of volume
fraction $\phi$. Such measurements are usually difficult \cite{Chien1966} because of possible slip, sedimentation and evaporation \cite{fung1996,Picart1998a}. Care has been taken
to avoid such problems, therefore only shear rates larger than $10^{-3} s^{-1}$ (lowest value) are considered. The empirical Herschel--Bulkley 
model (involving a yield stress) is then used when sufficient data points are available. The fits are in satisfactory agreement with the data which 
gives good confidence in the values of the yield stresses for $\phi \geq 0.42$. Another attempt has been made using Casson's model and gives similar 
data. The Bingham model seems to give less accurate values.

\begin{figure}
\centerline{
   \includegraphics[width=0.5\columnwidth]{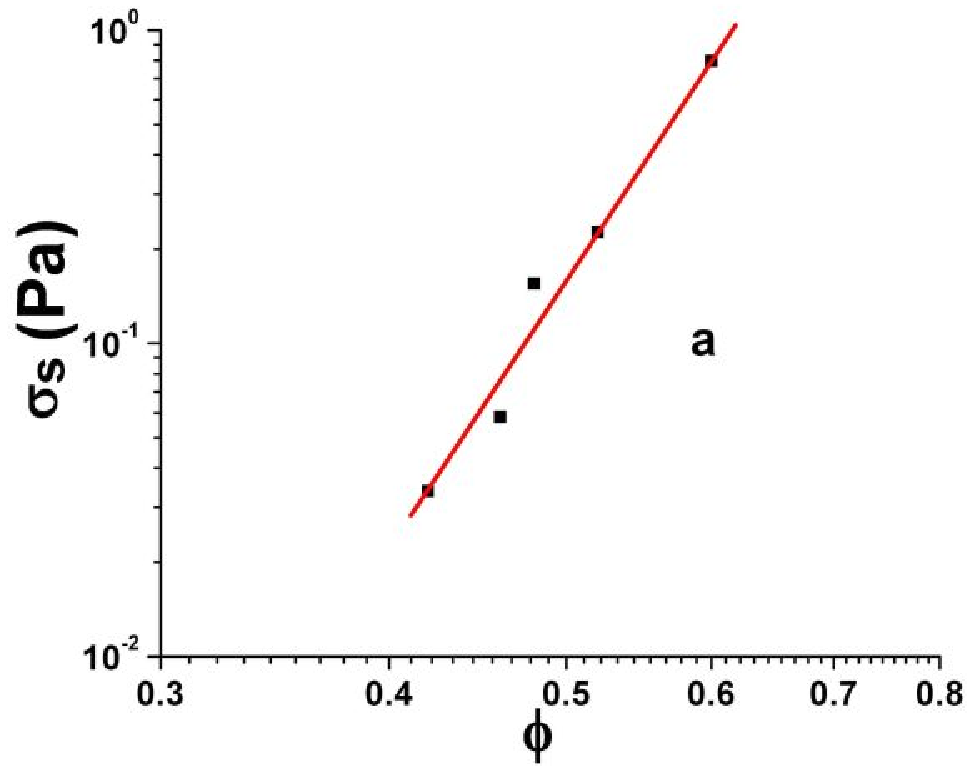}
   \includegraphics[width=0.5\columnwidth]{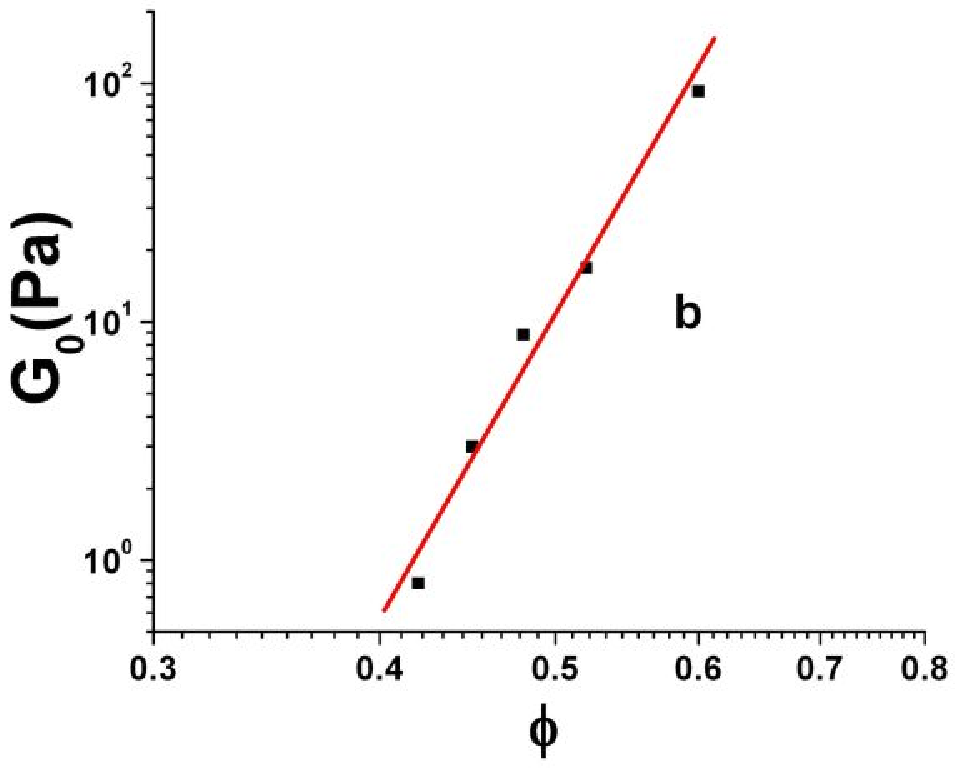}
}
\caption{\label{Fig:4} Yield stress $\sigma_{s}$ and Shear elastic modulus $G_{0}=G'(\omega=\,1 rad/s)$, vs. volume concentration $\phi$, log-scale.}
\end{figure}

The values of the yield stresses as a function of volume concentration are plotted in Fig.\ref{Fig:4}a, to be compared with the evolution of the 
shear plateau modulus $G_{0}$ (value of $G'$ at a typical angular frequency $\omega=1 \,rad/s$) in Fig.\ref{Fig:4}b. These two plots show power law 
dependences of the form  $\sigma_{s}\sim \phi^{m_{1}}$ and $G_{0}\sim \phi^{m_{2}}$ and remind previous results \cite{snabre1996} obtained in the case 
of the rheology of RBCs suspensions, at least for the yield stress $\sigma_{s}$. From Fig.\ref{Fig:4}a--b we find that $m_{1} \sim 8.4$ and
$m_{2} \sim 11.6$. These exponents are quite different from the ones obtained in the case of RBCs suspensions and this will be discussed below.

\begin{figure}
\includegraphics[width=1.0\columnwidth]{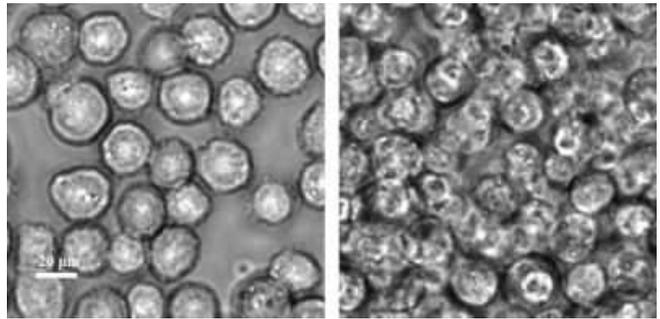}
\caption{\label{Fig:5} Phase contrast microscopy of CHO cell suspension : $10\%$ and $52\%$. Same
scale for both images.}
\end{figure}

{\it Modelling} -- As seen above, rheological modelling of such suspensions should therefore predict shear--thinning behavior, as well as yield 
stress properties at low shear rates $\dot \gamma \rightarrow 0$ and a concentration dependence of $\sigma_{s}$ and $G_{0}$. In addition, cell 
suspensions correspond to aggregated systems (see Fig.\ref{Fig:5}). Under flow, their structure is 
based on the persistent remodelling of the cells with respect to each other as they exhibit deformations, rotations, possible rolling and/or separation. 
During such events, cells may form clusters of size $R_{f}$ to be compared with the cell size $a$ (radius). 
The formation and destruction of cell clusters is the major ingredient to understand the rheological properties of the cell system, in order to explain
our previous data.
For example, when sheared under stress $\sigma$, clusters break into smaller ones, leading to shear--thinning effects. On the other hand, the 
possible encounter of clusters leads to the formation of larger structures, increasing the viscosity. 
Using a series of clich\'es like in Fig.\ref{Fig:5}, we measure the size $R_{f}$ of clusters at rest, found to vary as a power--law
of the type $\frac{R_{f}}{a}\sim N^{1/D}$ \cite{degennes1979}, where $N$ is the number of cells in a cluster and $D$ the fractal dimension. For our case, we find that 
$D \sim 1.47$ for 2D images. Thus, in three dimensions, we expect a fractal dimension of the order $D\sim 2$ \cite{kolb1984}. This number is similar to 
the ones found for RBCs suspensions, although the scaling exponents for yield stresses are quite different.

In the semi--empirical model proposed by Snabre and Mills \cite{snabre1996,snabre1999}, the formation and dissociation of clusters under flow is taken into 
account. A change in $R_{f}$ as a function of the applied shear stress is assumed : $\frac{R_{f}}{a}=1+(\frac{\sigma^{*}}{\sigma})^{m}$, 
where $m$ is a dimensionless parameter. $\sigma^{*}$ is a critical stress related to the interfacial adhesion between cells : $\sigma^{*}=\Gamma/a$, 
and $\Gamma$ is the cell adhesion free energy.
Using the concept of effective medium with volume fraction $\phi_{A}=\phi \, (\frac{R_{f}}{a})^{3-D}$, one assumes an effective viscosity 
$\eta(\sigma) = \eta_{0} \frac{1-\phi_{A}}{(1-\frac{\phi_{A}}{\phi_{0}})^{2}}$, and obtains the constitutive equation \cite{constit} which contains
the yield stress given by $\sigma_{s}\sim \sigma^{*} (\frac{\phi}{\phi_{0}})^{\frac{1}{m(3-D)}}$.
The last parameter to be used in the formula, $\phi_{0}$, is the maximum packing concentration found previously. 

We use the previous model to explain our experimental data. The exponent $m_{1}=8.4$ found for the yield stress $\sigma_{s}$ is plugged into
the previous scaling law for determination of the parameter $m=0.078$. 
This is smaller than the values of $m$ found for RBC suspensions (typically $m \sim 0.3$). This means that the size of clusters is not so sensitive 
to the applied stress, indeed one can consider that the cell aggregates are easier to form because of the round shape of the cells, in contrast 
with RBCs which need to bind in a very special way to form 'rouleaux', thus the stress has more effect on the latter ones.
We have obtained the value of the critical stress $\sigma^{*}=1.4 N/m^{2}$, and a corresponding value of $\Gamma = 1.4\,10^{-5} N/m$. 
This value of $\sigma^{*}$ is higher than the ones obtained for RBCs \cite{snabre1996} but the 
interfacial energy $\Gamma$ is in the range of the small values indicated for vesicles \cite{guttenberg2001}. This is in favor of the initial 
assumption that few adhesion molecules are involved in the region of contact between the cells. 

Finally, we postulate a similar relationship \cite{degennes1979} for the shear elastic modulus 
$G_{0}\sim  G^{*}(\frac{\phi}{\phi_{0}})^{\frac{1}{n(3-D)}}$, where $G^{*}$ is an effective elastic modulus, but we include an additional exponent $n$ 
to be determined. We come up with $n=0.056$ and $G^{*}=234 Pa$. This value of the reference modulus $G^{*}$, as explained in the concept of 
fractal exponents \cite{degennes1979}, is to be related to typical values for single cells. 
In particular, it gives a Young's elastic modulus of $E^{*}=702 Pa$ (assuming that the cell is incompressible) which is typical for adherent 
Wild Type CHO cells, of the order $0.5-1\,kPa$ as measured by AFM \cite{Canetta2005,Zhao2006b}.

To sum up, the system studied here provides unique features important for the rheology of biological suspensions and tissues. These concentrated cell 
suspensions behave as yield stress fluids (also called visco--plastic materials), for which a fractal approach has been used. Under shear, the fractal 
structure changes and can be modelled using a yield stress $\sigma_{s}$ and elasticity modulus $G_{0}$ related to the fractal dimension $D$. Two other parameters of 
interest have been introduced in the model: the adhesion energy $\Gamma$ found in the range of typical values for cell adhesion energies 
($\Gamma \sim 10^{-5} N/m$), and an effective elastic modulus ($E^{*}\sim 700\,Pa$) in agreement with previous microrheology experiments. 
We also found a similar behavior between the dynamic shear moduli $G'(\omega)$ and $G"(\omega)$ in this study (at large $\phi$), and the ones obtained from 
microrheological studies on single cells \cite{Fabry2001,Alcaraz2003,Hoffman2006} using various techniques. Both show slowly increasing dynamic 
moduli in terms of frequency, with the same relative positions. This idea probably deserves more attention and should be tested in the future, in particular 
further works may focus on the characterization of other cellular suspensions including cells with different elastic properties.

Finally, such a study can naturally lead to the understanding of biological tissues, by including stronger adhesion properties between the cells, or
by taking into account the addition of Extra-Cellular Matrix components.

{\it Acknowledgments} -- The authors wish to thank the EC Marie Curie Research Training Network MRTN-CT-2004-503661 
on "Modelling, mathematical methods and computer simulation of tumour growth and therapy" (http://calvino.polito.it/\~\,mcrtn/) 
for financial support, and C. Misbah for fruitful discussions about the manuscript.

\bibliographystyle{apsrev}
\bibliography{PRL}

\begin{thebibliography}{30}
\expandafter\ifx\csname natexlab\endcsname\relax\def\natexlab#1{#1}\fi
\expandafter\ifx\csname bibnamefont\endcsname\relax
  \def\bibnamefont#1{#1}\fi
\expandafter\ifx\csname bibfnamefont\endcsname\relax
  \def\bibfnamefont#1{#1}\fi
\expandafter\ifx\csname citenamefont\endcsname\relax
  \def\citenamefont#1{#1}\fi
\expandafter\ifx\csname url\endcsname\relax
  \def\url#1{\texttt{#1}}\fi
\expandafter\ifx\csname urlprefix\endcsname\relax\def\urlprefix{URL }\fi
\providecommand{\bibinfo}[2]{#2}
\providecommand{\eprint}[2][]{\url{#2}}

\bibitem[{\citenamefont{Larson}(1999)}]{Larson1999}
\bibinfo{author}{\bibfnamefont{R.~G.} \bibnamefont{Larson}},
  \emph{\bibinfo{title}{The structure and rheology of complex fluids}}
  (\bibinfo{publisher}{Oxford Univesity Press -- New-York},
  \bibinfo{year}{1999}).

\bibitem[{\citenamefont{Flatt and Bowen}(2006)}]{flatt2006}
\bibinfo{author}{\bibfnamefont{R.~J.} \bibnamefont{Flatt}} \bibnamefont{and}
  \bibinfo{author}{\bibfnamefont{P.}~\bibnamefont{Bowen}}, \bibinfo{journal}{J.
  Am. Ceram. Soc.} \textbf{\bibinfo{volume}{89}}, \bibinfo{pages}{1244}
  (\bibinfo{year}{2006}).

\bibitem[{\citenamefont{Alexandrou and Georgiou}(2007)}]{Alexandrou2007}
\bibinfo{author}{\bibfnamefont{A.~N.} \bibnamefont{Alexandrou}}
  \bibnamefont{and} \bibinfo{author}{\bibfnamefont{G.}~\bibnamefont{Georgiou}},
  \bibinfo{journal}{J. Non-Newtonian Fluid Mech.}
  \textbf{\bibinfo{volume}{142}}, \bibinfo{pages}{199} (\bibinfo{year}{2007}).

\bibitem[{\citenamefont{Brader et~al.}(2007)\citenamefont{Brader, Voigtmann,
  Cates, and Fuchs}}]{Brader2007}
\bibinfo{author}{\bibfnamefont{J.~M.} \bibnamefont{Brader}},
  \bibinfo{author}{\bibfnamefont{T.}~\bibnamefont{Voigtmann}},
  \bibinfo{author}{\bibfnamefont{M.~E.} \bibnamefont{Cates}}, \bibnamefont{and}
  \bibinfo{author}{\bibfnamefont{M.}~\bibnamefont{Fuchs}},
  \bibinfo{journal}{Phys. Rev. Lett.} \textbf{\bibinfo{volume}{98}},
  \bibinfo{pages}{058301} (\bibinfo{year}{2007}).

\bibitem[{\citenamefont{Coussot}(2005)}]{coussot2005}
\bibinfo{author}{\bibfnamefont{P.}~\bibnamefont{Coussot}},
  \emph{\bibinfo{title}{Rheometry of pastes, suspensions and granular
  materials}} (\bibinfo{publisher}{Wiley - New York}, \bibinfo{year}{2005}).

\bibitem[{\citenamefont{Laun}(1984)}]{Laun1984}
\bibinfo{author}{\bibfnamefont{H.~M.} \bibnamefont{Laun}},
  \bibinfo{journal}{Angew. Makromol. Chem.}
  \textbf{\bibinfo{volume}{124--125}}, \bibinfo{pages}{335}
  (\bibinfo{year}{1984}).

\bibitem[{\citenamefont{Chien et~al.}(1967{\natexlab{a}})\citenamefont{Chien,
  Usami, Dellenback, and Gregersen}}]{chien1967}
\bibinfo{author}{\bibfnamefont{S.}~\bibnamefont{Chien}},
  \bibinfo{author}{\bibfnamefont{S.}~\bibnamefont{Usami}},
  \bibinfo{author}{\bibfnamefont{R.~J.} \bibnamefont{Dellenback}},
  \bibnamefont{and} \bibinfo{author}{\bibfnamefont{M.~I.}
  \bibnamefont{Gregersen}}, \bibinfo{journal}{Science}
  \textbf{\bibinfo{volume}{157}}, \bibinfo{pages}{827}
  (\bibinfo{year}{1967}{\natexlab{a}}).

\bibitem[{\citenamefont{Chien et~al.}(1967{\natexlab{b}})\citenamefont{Chien,
  Usami, Dellenback, Gregersen, Nanninga, and Mason-Guest}}]{chien1967b}
\bibinfo{author}{\bibfnamefont{S.}~\bibnamefont{Chien}},
  \bibinfo{author}{\bibfnamefont{S.}~\bibnamefont{Usami}},
  \bibinfo{author}{\bibfnamefont{R.~J.} \bibnamefont{Dellenback}},
  \bibinfo{author}{\bibfnamefont{M.~I.} \bibnamefont{Gregersen}},
  \bibinfo{author}{\bibfnamefont{L.~B.} \bibnamefont{Nanninga}},
  \bibnamefont{and}
  \bibinfo{author}{\bibfnamefont{M.}~\bibnamefont{Mason-Guest}},
  \bibinfo{journal}{Science} \textbf{\bibinfo{volume}{157}},
  \bibinfo{pages}{829} (\bibinfo{year}{1967}{\natexlab{b}}).

\bibitem[{\citenamefont{Casson}(1959)}]{Casson1959}
\bibinfo{author}{\bibfnamefont{N.}~\bibnamefont{Casson}},
  \emph{\bibinfo{title}{A flow equation for pigment--oil suspensions of the
  printing ink type}} (\bibinfo{publisher}{Pergamon--London},
  \bibinfo{year}{1959}), chap.~\bibinfo{chapter}{5}.

\bibitem[{\citenamefont{Chien et~al.}(1966)\citenamefont{Chien, Usami, Taylor,
  Lundberg, and Gregersen}}]{Chien1966}
\bibinfo{author}{\bibfnamefont{S.}~\bibnamefont{Chien}},
  \bibinfo{author}{\bibfnamefont{S.}~\bibnamefont{Usami}},
  \bibinfo{author}{\bibfnamefont{H.~M.} \bibnamefont{Taylor}},
  \bibinfo{author}{\bibfnamefont{J.~L.} \bibnamefont{Lundberg}},
  \bibnamefont{and} \bibinfo{author}{\bibfnamefont{M.~I.}
  \bibnamefont{Gregersen}}, \bibinfo{journal}{J. Appl. Physiology}
  \textbf{\bibinfo{volume}{21}}, \bibinfo{pages}{81} (\bibinfo{year}{1966}).

\bibitem[{\citenamefont{Picart et~al.}(1998)\citenamefont{Picart, Piau,
  Galliard, and Carpentier}}]{Picart1998a}
\bibinfo{author}{\bibfnamefont{C.}~\bibnamefont{Picart}},
  \bibinfo{author}{\bibfnamefont{J.~M.} \bibnamefont{Piau}},
  \bibinfo{author}{\bibfnamefont{H.}~\bibnamefont{Galliard}}, \bibnamefont{and}
  \bibinfo{author}{\bibfnamefont{P.}~\bibnamefont{Carpentier}},
  \bibinfo{journal}{J. Rheol.} \textbf{\bibinfo{volume}{42}},
  \bibinfo{pages}{1} (\bibinfo{year}{1998}).

\bibitem[{\citenamefont{Snabre and Mills}(1996)}]{snabre1996}
\bibinfo{author}{\bibfnamefont{P.}~\bibnamefont{Snabre}} \bibnamefont{and}
  \bibinfo{author}{\bibfnamefont{P.}~\bibnamefont{Mills}}, \bibinfo{journal}{J.
  Phys. III. France} \textbf{\bibinfo{volume}{6}}, \bibinfo{pages}{1811}
  (\bibinfo{year}{1996}).

\bibitem[{\citenamefont{Snabre and Mills}(1999)}]{snabre1999}
\bibinfo{author}{\bibfnamefont{P.}~\bibnamefont{Snabre}} \bibnamefont{and}
  \bibinfo{author}{\bibfnamefont{P.}~\bibnamefont{Mills}},
  \bibinfo{journal}{Colloids Surf. Physicochem. Eng. Aspects}
  \textbf{\bibinfo{volume}{152}}, \bibinfo{pages}{79} (\bibinfo{year}{1999}).

\bibitem[{\citenamefont{Drasdo and H{\"o}hme}(2005)}]{Drasdo2005}
\bibinfo{author}{\bibfnamefont{D.}~\bibnamefont{Drasdo}} \bibnamefont{and}
  \bibinfo{author}{\bibfnamefont{S.}~\bibnamefont{H{\"o}hme}},
  \bibinfo{journal}{Phys. Biol.} \textbf{\bibinfo{volume}{2}},
  \bibinfo{pages}{133} (\bibinfo{year}{2005}).

\bibitem[{\citenamefont{Galle et~al.}(2005)\citenamefont{Galle, Loeffler, and
  Drasdo}}]{galle2005}
\bibinfo{author}{\bibfnamefont{J.}~\bibnamefont{Galle}},
  \bibinfo{author}{\bibfnamefont{M.}~\bibnamefont{Loeffler}}, \bibnamefont{and}
  \bibinfo{author}{\bibfnamefont{D.}~\bibnamefont{Drasdo}},
  \bibinfo{journal}{Biophys. J.} \textbf{\bibinfo{volume}{88}},
  \bibinfo{pages}{62} (\bibinfo{year}{2005}).

\bibitem[{\citenamefont{Ambrosi and Preziosi}(2007)}]{Ambrosi2007}
\bibinfo{author}{\bibfnamefont{D.}~\bibnamefont{Ambrosi}} \bibnamefont{and}
  \bibinfo{author}{\bibfnamefont{L.}~\bibnamefont{Preziosi}}
  (\bibinfo{year}{2007}), \bibinfo{note}{unpublished}.

\bibitem[{\citenamefont{Verdier}(2003)}]{Verdier2003}
\bibinfo{author}{\bibfnamefont{C.}~\bibnamefont{Verdier}}, \bibinfo{journal}{J.
  Theor. Medicine} \textbf{\bibinfo{volume}{5}}, \bibinfo{pages}{67}
  (\bibinfo{year}{2003}).

\bibitem[{\citenamefont{Thurston}(1972)}]{Thurston1972}
\bibinfo{author}{\bibfnamefont{G.~B.} \bibnamefont{Thurston}},
  \bibinfo{journal}{Biophys. J.} \textbf{\bibinfo{volume}{12}},
  \bibinfo{pages}{1205} (\bibinfo{year}{1972}).

\bibitem[{\citenamefont{Fabry et~al.}(2001)\citenamefont{Fabry, Maksym, Butler,
  Glogauer, Navajas, and Fredberg}}]{Fabry2001}
\bibinfo{author}{\bibfnamefont{B.}~\bibnamefont{Fabry}},
  \bibinfo{author}{\bibfnamefont{G.~N.} \bibnamefont{Maksym}},
  \bibinfo{author}{\bibfnamefont{J.~P.} \bibnamefont{Butler}},
  \bibinfo{author}{\bibfnamefont{M.}~\bibnamefont{Glogauer}},
  \bibinfo{author}{\bibfnamefont{D.}~\bibnamefont{Navajas}}, \bibnamefont{and}
  \bibinfo{author}{\bibfnamefont{J.~J.} \bibnamefont{Fredberg}},
  \bibinfo{journal}{Phys. Rev. Lett.} \textbf{\bibinfo{volume}{87}},
  \bibinfo{pages}{148102} (\bibinfo{year}{2001}).

\bibitem[{\citenamefont{Alcaraz et~al.}(2003)\citenamefont{Alcaraz, Buscemi,
  Grabulosa, Trepat, Fabry, Farr{\'e}, and Navajas}}]{Alcaraz2003}
\bibinfo{author}{\bibfnamefont{J.}~\bibnamefont{Alcaraz}},
  \bibinfo{author}{\bibfnamefont{L.}~\bibnamefont{Buscemi}},
  \bibinfo{author}{\bibfnamefont{M.}~\bibnamefont{Grabulosa}},
  \bibinfo{author}{\bibfnamefont{X.}~\bibnamefont{Trepat}},
  \bibinfo{author}{\bibfnamefont{B.}~\bibnamefont{Fabry}},
  \bibinfo{author}{\bibfnamefont{R.}~\bibnamefont{Farr{\'e}}},
  \bibnamefont{and} \bibinfo{author}{\bibfnamefont{D.}~\bibnamefont{Navajas}},
  \bibinfo{journal}{Biophys. J.} \textbf{\bibinfo{volume}{84}},
  \bibinfo{pages}{2071} (\bibinfo{year}{2003}).

\bibitem[{\citenamefont{Hoffman et~al.}(2006)\citenamefont{Hoffman, Massiera,
  Citters, and Crocker}}]{Hoffman2006}
\bibinfo{author}{\bibfnamefont{B.~D.} \bibnamefont{Hoffman}},
  \bibinfo{author}{\bibfnamefont{G.}~\bibnamefont{Massiera}},
  \bibinfo{author}{\bibfnamefont{K.~M.~V.} \bibnamefont{Citters}},
  \bibnamefont{and} \bibinfo{author}{\bibfnamefont{J.~C.}
  \bibnamefont{Crocker}}, \bibinfo{journal}{Proc. Natl Acad. Sci. USA}
  \textbf{\bibinfo{volume}{103}}, \bibinfo{pages}{10259}
  (\bibinfo{year}{2006}).

\bibitem[{\citenamefont{Quemada}(1998)}]{quemada1998a}
\bibinfo{author}{\bibfnamefont{D.}~\bibnamefont{Quemada}},
  \bibinfo{journal}{Eur. Phys. J. AP} \textbf{\bibinfo{volume}{1}},
  \bibinfo{pages}{119} (\bibinfo{year}{1998}).

\bibitem[{\citenamefont{Krieger and Dougherty}(1959)}]{Krieger1959}
\bibinfo{author}{\bibfnamefont{I.~M.} \bibnamefont{Krieger}} \bibnamefont{and}
  \bibinfo{author}{\bibfnamefont{T.~J.} \bibnamefont{Dougherty}},
  \bibinfo{journal}{Trans. Soc. Rheology} \textbf{\bibinfo{volume}{3}},
  \bibinfo{pages}{137} (\bibinfo{year}{1959}).

\bibitem[{\citenamefont{Fung}(1996)}]{fung1996}
\bibinfo{author}{\bibfnamefont{Y.~C.} \bibnamefont{Fung}},
  \emph{\bibinfo{title}{Biomechanics. Mechanical properties of living tissues}}
  (\bibinfo{publisher}{Springer -- New York}, \bibinfo{year}{1996}).

\bibitem[{\citenamefont{de~Gennes}(1979)}]{degennes1979}
\bibinfo{author}{\bibfnamefont{P.~G.} \bibnamefont{de~Gennes}},
  \emph{\bibinfo{title}{Scaling concepts in polymer physics}}
  (\bibinfo{publisher}{Cornell University Press}, \bibinfo{year}{1979}).

\bibitem[{\citenamefont{Kolb and Jullien}(1984)}]{kolb1984}
\bibinfo{author}{\bibfnamefont{M.}~\bibnamefont{Kolb}} \bibnamefont{and}
  \bibinfo{author}{\bibfnamefont{R.}~\bibnamefont{Jullien}},
  \bibinfo{journal}{J. Physique Lett.} \textbf{\bibinfo{volume}{45}},
  \bibinfo{pages}{977} (\bibinfo{year}{1984}).

\bibitem[{con()}]{constit}
\bibinfo{note}{The 1D--constitutive equation is given by
  $\eta(\sigma)=\frac{\sigma}{\dot\gamma}=\eta_{0}\frac{1-\phi(1+(\frac{\sigma%
_{*}}{\sigma})^{m})^{3-D}}
  {(1-\frac{\phi}{\phi_{0}}(1+(\frac{\sigma_{*}}{\sigma})^{m})^{3-D})^2}$ which
  is implicit for $\sigma$ but provides an explicit relation for $\dot \gamma$
  in terms of $\sigma$. Letting $\dot \gamma \rightarrow 0$ allows to determine
  the yield stress $\sigma_{s} $.}

\bibitem[{\citenamefont{Guttenberg et~al.}(2001)\citenamefont{Guttenberg, Lorz,
  Sackmann, and Boulbitch}}]{guttenberg2001}
\bibinfo{author}{\bibfnamefont{Z.}~\bibnamefont{Guttenberg}},
  \bibinfo{author}{\bibfnamefont{B.}~\bibnamefont{Lorz}},
  \bibinfo{author}{\bibfnamefont{E.}~\bibnamefont{Sackmann}}, \bibnamefont{and}
  \bibinfo{author}{\bibfnamefont{A.}~\bibnamefont{Boulbitch}},
  \bibinfo{journal}{Europhys. Lett.} \textbf{\bibinfo{volume}{54}},
  \bibinfo{pages}{826} (\bibinfo{year}{2001}).

\bibitem[{\citenamefont{Canetta et~al.}(2005)\citenamefont{Canetta, Duperray,
  Leyrat, and Verdier}}]{Canetta2005}
\bibinfo{author}{\bibfnamefont{E.}~\bibnamefont{Canetta}},
  \bibinfo{author}{\bibfnamefont{A.}~\bibnamefont{Duperray}},
  \bibinfo{author}{\bibfnamefont{A.}~\bibnamefont{Leyrat}}, \bibnamefont{and}
  \bibinfo{author}{\bibfnamefont{C.}~\bibnamefont{Verdier}},
  \bibinfo{journal}{Biorheology} \textbf{\bibinfo{volume}{42}},
  \bibinfo{pages}{321} (\bibinfo{year}{2005}).

\bibitem[{\citenamefont{Zhao et~al.}(2006)\citenamefont{Zhao, Srinivasan,
  Burgess, and Huey}}]{Zhao2006b}
\bibinfo{author}{\bibfnamefont{M.}~\bibnamefont{Zhao}},
  \bibinfo{author}{\bibfnamefont{C.}~\bibnamefont{Srinivasan}},
  \bibinfo{author}{\bibfnamefont{D.~J.} \bibnamefont{Burgess}},
  \bibnamefont{and} \bibinfo{author}{\bibfnamefont{B.~D.} \bibnamefont{Huey}},
  \bibinfo{journal}{J. Mater. Res.} \textbf{\bibinfo{volume}{21}},
  \bibinfo{pages}{1906} (\bibinfo{year}{2006}).

\end{thebibliography}

\end{document}